\documentclass{emulateapj}
\shorttitle{Molecular line observations of the Tornado}
\shortauthors{Sakai et al.}

\def\kms{\hbox{km s$^{-1}$}}

\begin{document}
\title{Millimeter-wave Molecular Line Observations of the Tornado Nebula}

\author{D. Sakai$^{1,2}$, T. Oka$^2$, K. Tanaka$^2$, S. Matsumura$^2$, K. Miura$^2$, and S. Takekawa$^2$}
\affil{$^1$Department of Astronomy, Graduate School of Science, The University of Tokyo, 7-3-1 Hongo, Bunkyo-ku, Tokyo 113-0033, Japan \\
$^2$Department of Physics, Institute of Science and Technology, Keio University, 3-14-1 Hiyoshi, Yokohama, Kanagawa 223-8522, Japan}

\email{sakai.daisuke@nao.ac.jp}

\begin{abstract}
We report the results of millimeter-wave molecular line observations of the Tornado Nebula (G357.7--0.1), which is a bright radio source behind the Galactic Center region.  A $15'\times 15'$ area was mapped in the {\it J}=1--0 lines of CO, $^{13}$CO, and HCO$^{+}$ with the Nobeyama Radio Observatory 45-m telescope.  The VLA archival data of OH at 1720$\,$MHz were also reanalyzed.  We found two molecular clouds with separate velocities, $V_{\rm LSR}=-14\,{\rm km\ s^{-1}}$ and $+5\,{\rm km\ s^{-1}}$.  These clouds show rough spatial anti-correlation.   Both clouds are associated with OH 1720 MHz emissions in the area overlapping with the Tornado Nebula.  The spatial and velocity coincidence indicates violent interaction between the clouds and the Tornado nebula.  Modestly excited gas prefers the position of the Tornado ``head'' in the $-14\,{\rm km\ s^{-1}}$ cloud, also suggesting the interaction.  Virial analysis shows that the $+5\,{\rm km\ s^{-1}}$ cloud is more tightly bound by self-gravity than the $-14\,{\rm km\ s^{-1}}$ cloud.  We propose a formation scenario for the Tornado Nebula; the $+5\,{\rm km\ s^{-1}}$ cloud collided into the $-14\,{\rm km\ s^{-1}}$ cloud, generating a high-density layer behind the shock front, which activates a putative compact object by Bondi-Hoyle-Lyttleton accretion to eject a pair of bipolar jets.
\end{abstract}

\keywords{ISM: individual (G357.7$-$0.1) --- ISM: clouds --- ISM: jets and outflows}

\section{Introduction}
The Tornado Nebula (G357.7--0.1) is a peculiar radio source toward the Galactic Center region.  Its name arises from its unusual axially symmetric, elongated morphology (Shaver et al. 1985; Becker $\&$ Helfand 1985; Helfand $\&$ Becker 1985).  The radio emissions are apparently of non-thermal origin with a spectral index of $-0.6$ (Dickel et al. 1973; Slee $\&$ Dulk 1974).  Spatially, the Tornado Nebula consists of two parts; the ``head" is a bright part in the northwest, and the ``tail" is a faint part in the southeast.  A compact source called the ``eye,'' which is located approximately 30\arcsec\ north from the ``head,'' appears to be a foreground object not associated with the Tornado.

The origin of the Tornado Nebula has long been unknown. A number of scenarios have been proposed, ranging from a Galactic object to an extragalactic object.  Some scenarios are based on the head-tail structure of the Tornado,  in which the activity center is located in the ``head" (Weiler et al. 1980; Miley 1980; Shull et al. 1989).  Others are based on the bipolar structure; the driving source is centered at the midpoint of the nebula (Caswell et al. 1989; Manchester 1987).  
From ${\rm H_I}$ absorption measurements, the distance to the Tornado Nebula is constrained to be $ > 6\,{\rm kpc}$ (Radhakrishnan et al. 1972).
The detection of OH 1720$\,$MHz maser emissions at the northern edge of the ``head" indicates the association of a C-type shock, and its velocity leads a kinematical distance of 11.8$\,$kpc, placing the Tornado Nebula in the category of Galactic objects (Frail et al. 1996).

Recently, X-ray observations with the Suzaku satellite detected twin clumps of thermal plasma at both ends of the Tornado Nebula (Sawada et al. 2011).   The temperatures of these clumps are similar, $0.73_{-0.15}^{+0.15}\,{\rm keV}$ and $0.59_{-0.15}^{+0.18}\,{\rm keV}$. 
The distances to these X-ray plasma clumps estimated by the absorption column densities are $10.0_{-1.4}^{+1.8}$ and $11.2_{-2.3}^{+2.7}\,{\rm kpc}$, respectively.  In addition, they also found that molecular clouds are associated with each plasma clump, suggesting that the clumps have been formed by bipolar jets stopped by the molecular clouds.
The putative bipolar jets might have been generated by a compact object that was previously active and has been dimmed for at least 40--50$\,$years.

In this paper, we report the results of new millimeter-wave molecular line observations  and reanalysis of the VLA  archival data of the OH 1720$\,$MHz line toward the Tornado Nebula. On the basis of these new data sets, we discuss the formation mechanism of the Tornado Nebula along with the bipolar jet scenario.

\begin{figure*}[tbhp]
\epsscale{1}
\plotone{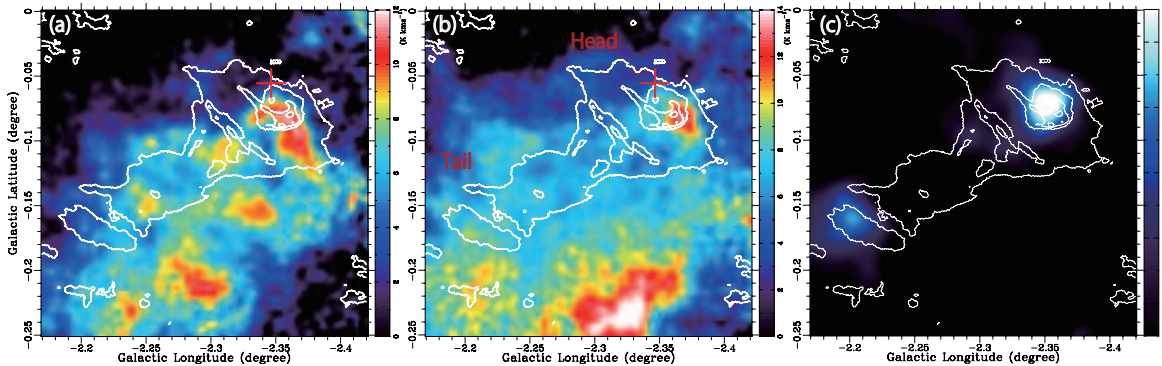}
\caption{(a) Velocity-integrated map of $^{13}$CO {\it J}=1--0 emission from $-15\,{\rm km\ s^{-1}}$ to $-13\,{\rm km\ s^{-1}}$. White contours represent the $1.58\,{\rm GHz}$ radio continuum map which has been published by Shaver et al. (1985), Becker and Helfand (1985), and Helfand and Becker (1985). The red cross denotes the position of the OH 1720$\,$MHz maser spot. (b) Velocity-integrated map of $^{12}$CO {\it J}=1--0 emissions over the same velocity range as in (a). (c) Smoothed broad-band X-ray image of the Tornado Nebula observed with the Suzaku satellite (Sawada et al. 2011). The 1.58$\,$GHz radio continuum contours are superposed. The band range displayed is 2.0--$5.0\,$keV.}
\end{figure*}

\section{Observations and Analyses}

\subsection{Nobeyama Radio Observatory 45-m observations}

We made observations of CO  {\it J}=1--0 (115.271204$\,$GHz), $^{13}$CO  {\it J}=1--0 (110.201353$\,$GHz), C$^{18}$O  {\it J}=1--0 (109.782182$\,$GHz), HCO$^+$  {\it J}=1--0 (89.188568$\,$GHz), and N$_2$H$^+$ {\it J} = 1--0 (93.173777$\,$GHz) of the Tornado Nebula with the Nobeyama Radio Observatory 45-m telescope from March 1 to 10 in 2013.
The typical system noise temperature during these observations was in the range of 200--400$\,$K.
All observations were made with the focal-plane array SIS receiver, BEARS (Sunada et al. 2000), in the on-the-fly (OTF) mapping mode (Sawada et al. 2008).
A $\sim 15' \times 15'$ area centered at $(l,b) = (-2^{\circ}\hspace{-3.5pt}.275,-0^{\circ}\hspace{-3.5pt}.125)$ was mapped to cover the full extent of the Tornado Nebula.
The reference position was $(l,b)=(+0^{\circ},-1^{\circ})$.
The antenna temperature was obtained by the standard chopper-wheel technique (Kutner \& Ulich 1981). The telescope pointing was corrected every hour by observing the SiO maser source VX Sgr and maintained at $\leq 3\arcsec$. The 25 digital auto-correlators were used as spectrometers in the wide-band mode, which has a 512$\,$MHz coverage (1400$\,{\rm km\ s^{-1}}$ at 110$\,$GHz) and a 500$\,$kHz resolution (1.4$\,{\rm km\ s^{-1}}$).
We scaled the antenna temperature by multiplying it by $1/\eta_{\rm MB}$ to obtain the main beam temperature, $T_{\rm MB}$. We used $\eta_{\rm MB} = 0.45\pm 0.04$ for 110$\,$GHz.

The obtained data were reduced by using the NOSTAR reduction package.
All the data were resampled onto a $7\arcsec\hspace{-3.5pt}.5 \times 7\arcsec\hspace{-3.5pt}.5\times 1\,{\rm km\ s^{-1}}$ grid to obtain the final maps. The rms noise level, $\Delta T_{\rm MB}$, of the map was 0.66$\,$K.
Both CO lines were detected in almost the whole area.
HCO$^+$ was detected only in some limited regions.
N$_2$H$^+$ emissions were not detected in these observations.

\subsection{VLA archival data analyses}

We obtained OH 1720$\,$MHz data from the archives of the Very Large Array (VLA) of the National Radio Astronomy Observatory (NRAO).
The observations of the Tornado Nebula had been conducted on June 24 and 18, 2000, in the C- and D-array configurations.
The synthesized beam size was $40\arcsec \times 23\arcsec$.

The OH 1720$\,$MHz data were reduced with Common Astronomy Software Applications (CASA) developed by the NRAO.
We used CLEAN as a deconvolution algorithm.
The parameters for CLEAN were set as $\gamma=0.1$, and $I_{\rm th} = 250\,{\rm mJy}$, where $\gamma$ is the gain factor, and $I_{\rm th}$ is the threshold intensity.
The grid spacing of the map was $15\arcsec \times 15\arcsec \times 1\,{\rm km\ s^{-1}}$.
The rms noise level of the map was 98 ${\rm mJy/beam}$.

\section{Results}

\subsection{CO distributions at $V_{\rm LSR}=-14\,{\rm km\ s^{-1}}$}

Figure 1 shows the distributions of $^{13}$CO and $^{12}$CO {\it J}=1--0 emissions toward the Tornado Nebula near the velocity of $-14\,{\rm km\ s^{-1}}$, at which a compact OH 1720$\,$MHz maser spot had been detected (Frail et al. 1996).
We detected CO lines from more than $\sim2/3$ of the mapping area. 

The widespread CO distribution consists of several clumps.
The most intense emission originates from the center of the ``head" of the Tornado Nebula at $(l,b)=(-2.\hspace{-3.5pt}^{\circ}35,-0.\hspace{-3.5pt}^{\circ}10)$.
This corresponds to ``MC1" in Sawada et al. (2011).
Both the position and velocity of this component coincide with those of the OH $1720\,$MHz maser spot (Frail et al. 1996).
Thus, this component should be associated with the Tornado Nebula.
While this component appears as a clump with a $\sim 0^{\circ}\hspace{-3.5 pt}.02$ diameter in the $^{12}$CO map, it has an arc-shaped appearance with a $\sim 0^{\circ}\hspace{-3.5 pt}.04$ diameter in $^{13}$CO.

A fluffy component in the bottom-left of the $^{12}$CO map corresponds to ``MC2" (Sawada et al. 2011).
However, there are no strong emissions in the same area of the $^{13}$CO map (Fig. 1{\it a}).

In addition to ``MC1", two $^{13}$CO components at $(l, b)=(-2^{\circ}\hspace{-3.5pt}.25, -0^{\circ}\hspace{-3.5pt}.15)$ and ($-2^{\circ}\hspace{-3.5pt}.30$, $-0^{\circ}\hspace{-3.5pt}.11$) overlap with the Tornado Nebula on the plane of the sky.
Those three components are aligned along the main-axis of the Tornado Nebula.
The other $^{13}$CO components, at ($l$, $b$)=($-2^{\circ}\hspace{-3.5pt}.23$, $-0^{\circ}\hspace{-3.5pt}.23$), ($-2^{\circ}\hspace{-3.5pt}.3$, $-0^{\circ}\hspace{-3.5pt}.22$), and ($-2^{\circ}\hspace{-3.5pt}.34$, $-0^{\circ}\hspace{-3.5pt}.16$), do not overlap with the Tornado Nebula.

\subsection{Velocity channel maps}

\begin{figure*}[tbhp]
\epsscale{1}
\plotone{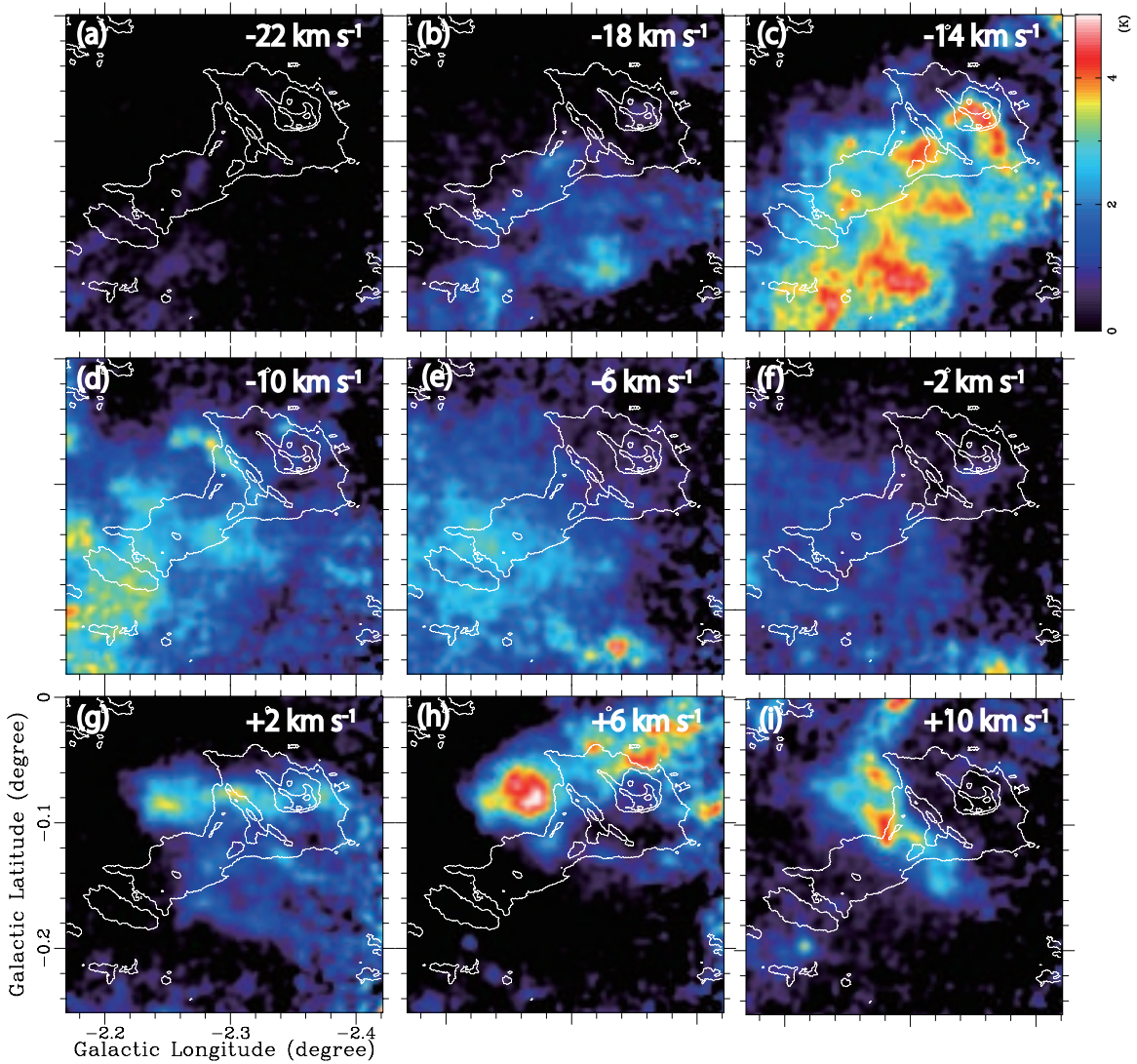}
\caption{Velocity channel maps of $^{13}$CO {\it J}=1--0 from $-22\,{\rm km \ s^{-1}}$ to $+10\,{\rm km \ s^{-1}}$.  The velocity range increases from ({\it a}) to ({\it i}) and the velocity width of each channel is $4\,{\rm km\ s^{-1}}$. White contours represent a 1.58$\,$GHz radio continuum map.}
\end{figure*}

The velocity channel maps of $^{13}$CO {\it J}=1--0 are shown in Fig. 2.
Those maps cover a range from $-22\,{\rm km\ s^{-1}}$ to $+10\,{\rm km \ s^{-1}}$ with a $4\,{\rm km\ s^{-1}}$ interval.
The $^{13}$CO components in Fig. 1{\it a} appears in velocities between $-18\,{\rm km\ s^{-1}}$ and $-10\,{\rm km\ s^{-1}}$ (Fig.2{\it b--d}).

In addition to this velocity component at -14 km/s, toward which the OH maser spot was reported, we also see several $^{13}$CO clumps in higher velocity channels.
 A small clump, which did not appear in Fig. 1{\it a}--{\it b}, is detected in the bottom of Fig. 2{\it e}.
In positive LSR velocities, there are two $^{13}$CO components adjacent to the Tornado Nebula.
One is in the left side, and another is detected in the upper side of the Tornado Nebula.
These positive-velocity components seem to be spatially anti-correlated with the Tornado Nebula and with the negative-velocity components.

\subsection{Longitude-velocity map}

\begin{figure}[htbp]
\epsscale{1}
\plotone{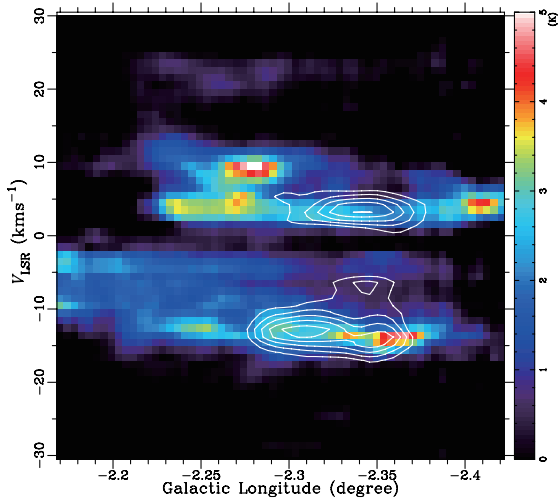}
\caption{Longitude-velocity map of $^{13}$CO {\it J}=1--0 emission integrated over latitudes from $b=-0^{\circ}\hspace{-3.5pt}.12$ to $-0^{\circ}\hspace{-3.5pt}.06$. White contours represent longitude-velocity distribution of OH 1720$\,$MHz emissions over the same latitude range as $^{13}$CO. Contours are set at $0.1\,{\rm Jy\ beam^{-1}}$ intervals from $1.2\,{\rm Jy\ beam^{-1}}$.}
\end{figure}

Figure 3 shows the longitude-velocity map of $^{13}$CO {\it J}=1--0 emissions integrated over latitudes from $b=-0^{\circ}\hspace{-3.5pt}.12$ to $-0^{\circ}\hspace{-3.5pt}.06$.
The same map of OH 1720$\,$MHz emissions is superposed.
This map indicates that the $^{13}$CO emission components toward the Tornado Nebula are roughly classified into two velocity groups.
Each group of emission components seems to form a loosely bound molecular cloud.
Velocity ranges of these clouds are $-17\,{\rm km\ s^{-1}}$ to $0\,{\rm km\ s^{-1}}$ and $0\,{\rm km\ s^{-1}}$ to $+15\,{\rm km\ s^{-1}}$, whereas the most prominent components appear at $-14\,{\rm km\ s^{-1}}$ and $+5\,{\rm km\ s^{-1}}$, respectively.
Thus, we refer to the former as the $-14\,{\rm km\ s^{-1}}$ cloud and to the latter as the $+5\,{\rm km\ s^{-1}}$ cloud.
The OH 1720$\,$MHz emissions are associated with the primary parts of those clouds.
The extended OH emissions at $-14\,{\rm km\ s^{-1}}$ have already been reported in Yusef-Zadeh et al. (1999).

\subsection{$-14\,{\rm km\ s^{-1}}$ and $+5\,{\rm km\ s^{-1}}$ clouds}

 \begin{figure*}[thbp]
\epsscale{1}
\plotone{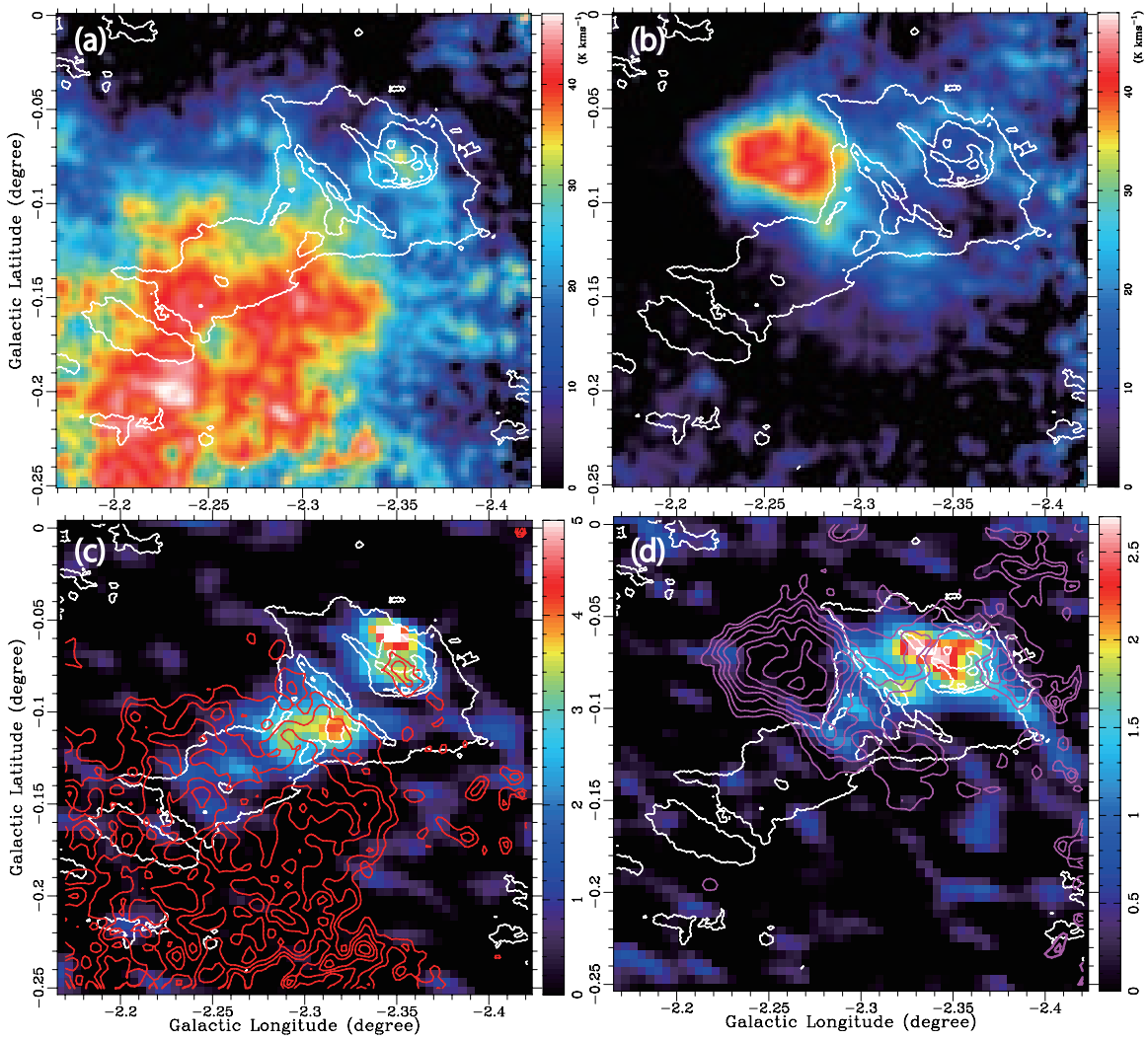}
\caption{Maps of velocity-integrated molecular line emissions. (a) $^{13}$CO {\it J}=1--0 line. The velocity range for the integration is from $V_{\rm LSR} = -17\,{\rm km\ s^{-1}}$ to $0\,{\rm km\ s^{-1}}$. (b) $^{13}$CO {\it J}=1--0 line. The velocity range is from $V_{\rm LSR}=0\,{\rm km\ s^{-1}}$ to $+15\,{\rm km\ s^{-1}}$. (c) OH 1720$\,$MHz line. The velocity range is from $V_{\rm LSR}=-17\,{\rm km\ s^{-1}}$ to $0\,{\rm km\ s^{-1}}$. Red contours represent velocity-integrated $^{13}$CO line emissions in the same range as in (a). The contours are set at $5\,{\rm K\ km\ s^{-1}}$ intervals from $25\,{\rm K\ {\rm km\ s^{-1}}}$. (d) OH 1720$\,$MHz line. The velocity range is from $V_{\rm LSR}=0\,{\rm km\ s^{-1}}$ to $15\,{\rm km\ s^{-1}}$. Purple contours represent velocity-integrated $^{13}$CO line emissions in the same range as in (b). The contours are set at $5\,{\rm K\ km\ s^{-1}}$ intervals from $15\,{\rm K\ {\rm km\ s^{-1}}}$. }
\end{figure*}

Figures 4{\it a} and {\it b} show velocity-integrated maps of $^{13}$CO {\it J}=1--0 lines for the $-14\,{\rm km\ s^{-1}}$ and $+5\,{\rm km\ s^{-1}}$ clouds.
Each map is integrated over $-17\,{\rm km\ s^{-1}}$ to $0\,{\rm km\ s^{-1}}$ and $0\,{\rm km\ s^{-1}}$ to $+15\,{\rm km\ s^{-1}}$, respectively.
Integrated intensity maps of the OH 1720 MHz for the same velocity ranges are shown in Fig. 4{\it c} and {\it d}. In these two maps, the $^{13}$CO {\it J}=1--0 maps are superposed.

The $-14\,{\rm km\ s^{-1}}$ cloud has a fluffy appearance, whereas the $+5\,{\rm km\ s^{-1}}$ cloud is particulaly compact.
HCO$^+$ {\it J}=1--0 line emissions are detected only in the region of the $+5\,{\rm km\ s^{-1}}$ cloud and the point at which the OH 1720$\,$MHz maser was detected.
These $^{13}$CO clouds show clear spatial anti-correlation with each other.

The OH 1720 MHz emissions from the both -14 km/s and +5 km/s clouds seem spatially well correlated to the Tornado nebula, especially to the ``head'' part of the nebula.  
It is known that OH 1720$\,$MHz emissions are enhanced in a region where a C-type shock has just passed (Elitzur 1976; Lockett et al. 1999).
Figures 4{\it c} and {\it d} show that the OH 1720 MHz emissions are bright in the overlapping regions of the $^{13}$CO and the Tornado for the both velocity components.  
Hence, these OH emissions are likely to trace the region violently interacting with the Tornado Nebula.

The $+5\,{\rm km\ s^{-1}}$ component corresponds to the $+5\,{\rm km\ s^{-1}}$ $^{13}$CO cloud, whereas the $-12\,{\rm km\ s^{-1}}$ component is not strongly correlated with the $-14\,{\rm km\ s^{-1}}$ $^{13}$CO cloud.
Although another velocity component at $V_{\rm LSR} = -34\,{\rm km\ s^{-1}}$ is detected, this component has a CO counterpart, which may belong to the Norma Arm (Sofue et al. 2006).

\section{Discussion}
\subsection{Excitation temperature distribution}
Here, we refer to the excitation temperature to investigate the physical relation between the Tornado Nebula and adjacent molecular clouds.  The CO excitation temperature ($T_{\rm ex}$) was calculated by the following procedure (Oka et al. 1998).

The observed radiation temperature can be expressed as
\begin{equation}
T_{\rm R}^* = f[J(T_{\rm ex}) - J(T_{\rm CMB})][1 - \exp(-\tau)]
\end{equation}
where $f$ is the beam filling factor of the emitting region, $J(T) \equiv (h\nu/k)[\exp (h\nu/kT) - 1]$, $T_{\rm ex}$ is the excitation temperature between the upper and the lower levels of the transition, $T_{\rm CMB}$ is the cosmic microwave background temperature, and $\tau$ is the optical depth of the line.
The optical depth of the line can be estimated from the ratio of $^{12}$CO and $^{13}$CO intensities,
\begin{equation}
\frac{T_{\rm R}^{*12}}{T_{\rm R}^{*13}} = \frac{f_{12}}{f_{13}} \frac{J(T_{\rm ex}^{12}) - J(T_{\rm CMB})}{J(T_{\rm ex}^{13}) - J(T_{\rm CMB})} \frac{1 - \exp (-\tau_{12})}{1 - \exp(-\tau_{13})}
\end{equation}
where $\tau_{13} = \tau_{12}/([^{12}$CO$]/[^{13}$CO$])$.
We assumed $[^{12}$CO$]/[^{13}$CO$] = 60$ (Langer $\&$ Penzias 1990), $f_{12}=f_{13}$, and $T_{\rm ex}^{12} = T_{\rm ex}^{13}$.
The excitation temperature can be estimated from $^{12}$CO radiation temperature and optical depth by equation (1) assuming that the beam filling factor is unity ($f = 1$).

We calculated $T_{\rm ex}$ for data points where both $^{12}$CO and $^{13}$CO lines are detected over $1\sigma $ significance.  We obtained the $T_{\rm ex}$ data cube with $7\arcsec\hspace{-3.5pt}.5 \times 7\arcsec\hspace{-3.5pt}.5\times 1\,{\rm km\ s^{-1}}$ grid spacing.  $T_{\rm ex}$ was successfully calculated for 16\% of all data points. 
Calculated $T_{\rm ex}$ values are lower limits because the beam filling factor is usually lower than unity.

\begin{figure*}[phbt]
\epsscale{1}
\plotone{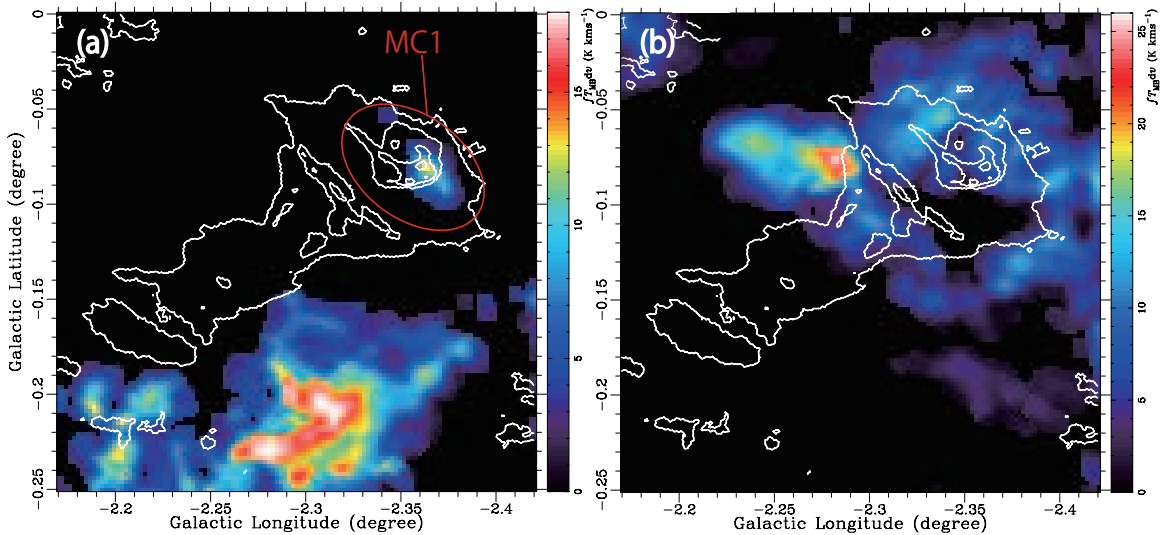}
\caption{Maps of velocity-integrated $^{13}$CO {\it J}=1--0 line emissions, with low-excitation temperature emissions omitted. (a) The velocity range for the integration is from $V_{\rm LSR} = -17\,{\rm km\ s^{-1}}$ to $0\,{\rm km\ s^{-1}}$. The threshold excitation temperature, $T_{\rm th}$, is set to 8$\,$K. (b) The velocity range for the integration is from $V_{\rm LSR} = 0\,{\rm km\ s^{-1}}$ to $+15\,{\rm km\ s^{-1}}$. The threshold excitation temperature is set to 7$\,$K.}
\end{figure*}

To illustrate the spatial distribution of high-$T_{\rm ex}$ gas, we created maps of $^{13}$CO {\it J}=1--0 emission by integrating the data with $T_{\rm ex}\geq T_{\rm th}$, where $T_{\rm th}$ is the threshold temperature.
Figure 5 shows the distributions of high-$T_{\rm ex}$ gas in the $-14\,{\rm km\ s^{-1}}$ and $+5\,{\rm km\ s^{-1}}$ clouds, with $T_{\rm th}=8\,$K and $7\,$K, respectively.  High-$T_{\rm ex}$ gas in the $-14\,{\rm km\ s^{-1}}$ cloud spreads in its southern half, while a spot pinpoints the radio continuum peak in the Tornado head.  This spot corresponds to MC1 in Sawada et al. (2011).  High-$T_{\rm ex}$ gas in the $+5\,{\rm km\ s^{-1}}$ cloud clearly surrounds the Tornado head.  These results suggest that high-$T_{\rm ex}$ gas is closely associated with the Tornado head, except for the southern half of the $-14\,{\rm km\ s^{-1}}$ cloud, which might have a different cause for high excitations.

\subsection{Properties of the $-14\,{\rm km\ s^{-1}}$ and $+5\,{\rm km\ s^{-1}}$ clouds}

To investigate the properties of the two $^{13}$CO clouds, we performed Virial analysis on both clouds.
The LTE masses of the clouds, $M_{\rm LTE}$, were calculated by 
assuming that the gas temperature is 10$\,$K 
, and the isotopic abundance ratio, [$^{12}$CO]/[$^{13}$CO], is 60 (Langer $\&$ Penzias 1990).
The derived LTE masses of the $-14\,{\rm km\ s^{-1}}$ and $+5\,{\rm km\ s^{-1}}$ clouds are $3.0 \times 10^{5}\,M_{\odot}$ and $5.5 \times 10^4\,M_{\odot}$, respectively.

On the other hand, the Virial mass, $M_{\rm VT}$, can be calculated as follows:
\begin{equation}
M_{\rm VT} = \frac{3R\sigma_v^2}{G}
\end{equation}
where $R$ is the size scale of the cloud, $\sigma_v$ is the velocity dispersion, and $G$ is the gravitational constant.
The size scales, $R$, are calculated by using the assumption that the distance to the Tornado Nebula is 11.8$\,$kpc (Frail et al. 1996; Sawada et al. 2011), which are 40$\,$pc and 10$\,$pc for the $-14\,{\rm km\ s^{-1}}$ cloud and $+5\,{\rm km\ s^{-1}}$ cloud, respectively.  The velocity dispersion, $\sigma_{v}$, are 3.4 \kms\ and 4.8 \kms\ , respectively.   The calculated $M_{\rm LTE}$ and $M_{\rm  VT}$ are listed in Table 1.  
\begin{table}
\begin{center}
\caption{The relationship between $M_{\rm LTE}$ and $M_{\rm VT}$}
\begin{tabular}{crr}
\tableline\tableline
Cloud & $-14\,{\rm km\ s^{-1}}$ & $+5\,{\rm km\ s^{-1}}$ \\
 Size & 40$\,$pc & 10$\,$pc \\
 $\sigma_v$ & $3.4\,{\rm km\ s^{-1}}$ & $4.8\,{\rm km\ s^{-1}}$\\
\tableline\tableline
$M_{\rm LTE}(M_{\odot})$ & $ 3.0\times 10^5$ & $ 5.5 \times 10^4$\\
$M_{\rm VT}(M_{\odot})$ & $6.4 \times 10^5$ & $8.0 \times 10^4$\\
$M_{\rm VT}/M_{\rm LTE}$ &  2.1 &  1.5 \\
\tableline
\end{tabular}
\end{center}
\end{table}

It is known that the ratio of $M_{\rm VT}/M_{\rm LTE}$ can be used as a measure of the gravitational instability of a cloud (Solomon et al. 1987; Oka et al. 2001).
The lower the ratio is, the more strongly the cloud is bounded.
Our analysis shows that the ratio of the $+5\,{\rm km\ s^{-1}}$ cloud is smaller than that of the $-14\,{\rm km\ s^{-1}}$ cloud; that is, the former is more strongly bounded than the latter.

\subsection{Interaction between the Tornado Nebula and the $^{13}$CO clouds}
From our observations and analyses, significant evidence for interaction between the Tornado Nebula and ambient molecular gas was obtained.  (1) Spatially extended OH 1720$\,$MHz emission was detected in the two $^{13}$CO clouds at same velocities; extended OH masers were often detected toward MCs interacting with C-type shock (Yusef-Zadeh et al. 1999). (2) Spatial anti-correlation between the Tornado and the two $^{13}$CO clouds was found. This is clearly seen in Fig. 4{\it b}, where the curvature of the western edge of the $+5$ \kms\ cloud seems to be in good accordance with the eastern edge of the Tornado. In addition, the core of the Tornado $(l,b)=(2^{\circ}\hspace{-3.5pt}.35, -0^{\circ}\hspace{-3.5pt}.07)$ is enclosed by a low-level $^{13}$CO emission. (3) Slight enhancement in $T_{\rm ex}$ in the area adjacent to or overlapped with the Tornado Nebula was found.  These clearly indicate physical contact between the Tornado Nebula and the $^{13}$CO clouds.  The violent interaction suggests that the Tornado Nebula is expanding.  This is consistent with the bipolar-jet driven scenario (e.g., Sawada et al. 2011). 
The detection of GeV $\gamma$-ray from the vicinity of the $+5$ \kms\ cloud (Castro et al. 2013) lends another support for the violent interaction between the cloud and the Tornado Nebula, since several SN/MC interacting systems have been detected in $\gamma$-ray (Pollack 1985; Hewitt et al. 2009). 

The interaction between the Tornado Nebula and the two $^{13}$CO clouds also indicates the close association of these clouds. The velocity difference of $\sim 20\,{\rm km\ s^{-1}}$ between the clouds indicates that these might have collided or will collide within the crossing time, $\sim 10^6$ years.  Rough spatial anti-correlation between the two $^{13}$CO clouds  (Fig. 2{\it d} and {\it h}) implies that the collision has occurred in the recent past.  If the $-14$ \kms\ cloud is moving along the Galactic rotation, the motion of the  $+5$ \kms\ cloud must deviate from that significantly.  The $-14$ \kms\  cloud and the Tornado are most likely on the Sagittarius arm, while the $+5$ \kms\ cloud may be on an elongated orbit along with the Galactic bar.   In fact, the location of the Tornado is in the vicinity of the far-side end of the Galactic bar, where the orbit intersection is expected.

\subsection{Formation Scenario of the Tornado Nebula}
Our observations and analyses are summarized as follows: 
\begin{itemize}
\item{Two molecular clouds, $-14$ \kms\ and $+5$ \kms\ clouds, are associated with the Tornado Nebula.}
\item{The Tornado Nebula is interacting with both of the $-14$ \kms\ and $+5$ \kms\ clouds.}
\item{The $-14$ \kms\ and $+5$ \kms\ clouds show a rough spatial anti-correlation, suggesting a physical contact.}
\end{itemize}

Taking these results and based on the bipolar jet model by Sawada et al. (2011), we propose a formation scenario for the Tornado Nebula: 
\begin{enumerate}
\item The $+5\,{\rm km\ s^{-1}}$ cloud collides with the $-14\,{\rm km\ s^{-1}}$ cloud, generating a shock in both clouds.  
\item A high-density layer behind the shock front in the  $-14\,{\rm km\ s^{-1}}$ cloud passes a massive compact object, which was assumed in Sawada et al. (2011).
\item The Bondi-Hoyle-Lyttleton accretion activates the putative compact object, which thereby ejects a pair of bipolar jets.  
\item The bipolar jets violently interact with the ambient molecular clouds, generating C-type shocks, and thereby form the Tornado Nebula and the twin plasma clumps.  
\end{enumerate}
Figure 6 shows a schematic view of this scenario.  

\begin{figure*}[htbp]
\epsscale{0.8}
\plotone{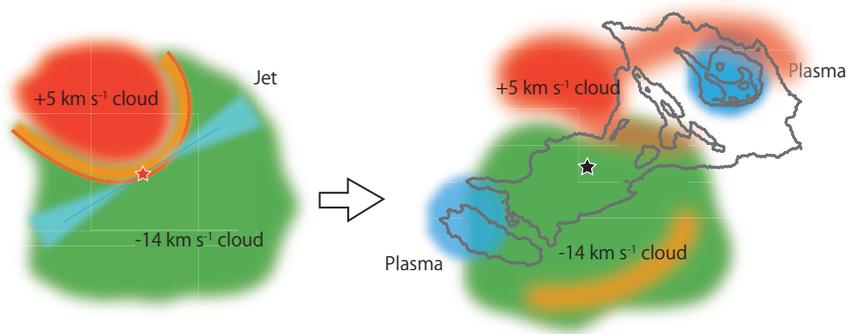}
\caption{Schematic view of a formation scenario of the Tornado Nebula.  The $+5\,{\rm km\ s^{-1}}$ cloud collides with the $-14\,{\rm km\, s^{-1}}$ cloud. A shock created by the collision activates a putative compact object. The left panel shows the configuration $\sim 10^6\,{\rm years}$ ago.  
The right panel shows the current configuration.}
\end{figure*}

The Virial analysis of the $^{13}$CO clouds supports the notion that the gravitationally bound $+5\,{\rm km\ s^{-1}}$ cloud plunged into the loosely bound $-14\,{\rm km\ s^{-1}}$ cloud.  The high-excitation area in the southern half of the $-14\, {\rm km\ s^{-1}}$ cloud could be a remnant of the shock front.  
The shock velocity should be less than the velocity difference between colliding clouds, $20\,{\rm km\ s^{-1}}$, which is in the range of a non-dissociative C-type shock (Wardle 1998; Lockett et al. 1999).
The mass accretion rate for the Bondi-Hoyle-Lyttleton (BHL) accretion, $\dot{M}_{\rm HL}$, is related to the critical impact parameter, $R_{\rm HL} = \frac{2GM}{v^2+\sigma^2}$, and can be expressed as
\begin{equation}
\dot{M}_{\rm HL} = \pi R_{\rm HL}^2 (v^2 + \sigma^2)^{1/2} \rho = \frac{4\pi G^2 M^2 \rho}{(v^2+\sigma^2)^{3/2}}
\end{equation}
where $M$ is the mass of a high-density object, $\rho$ is the mass density of the ambient interstellar medium, $v$ is the relative velocity, and $\sigma$ is the velocity dispersion (Edgar 2004).  
Assuming the standard accretion model onto a black hole, the BHL accretion with a time duration $\tau$ radiates
\begin{equation}
E_{\rm HL}=\frac{1}{12} \dot{M}_{\rm HL} c^2 \tau .
\end{equation}
If a cloud with a density of $n^0({\rm H}_2)$ and a depth of $l^0$ is compressed to $n({\rm H}_2)$ and $l$ by the cloud-to-cloud collision, and the compressed layer passes the black hole with the velocity $v$, the radiated energy becomes 
\begin{eqnarray}
E_{\rm HL} &=& 1.84\times 10^{50} \left( \frac{M}{30\,M_{\odot}} \right)^2 \left( \frac{n^0({\rm H}_2)}{10^2\,{\rm cm^{-3}}}\right) \left( \frac{l^0}{10\,{\rm pc}}\right)\nonumber \\
&& \times \left[ \frac{ v^2 + \sigma^2}{ (10\,{\rm km\ s^{-1}})^2 + (3\,{\rm km\ s^{-1}})^2} \right]^{-3/2}\nonumber \\
&&\times  \left(\frac{v}{10\,{\rm km\ s^{-1}}} \right)^{-1} \,{\rm erg}.
\end{eqnarray}
Thus the reasonable choice of parameters, $n^0({\rm H}_2)=10^2\,{\rm cm}^{-3}$, $l^0=10\,{\rm pc}$, $v=10$ \kms, $\sigma=3$ \kms, and the existence of a $M\sim 50$ $M_{\sun}$ black hole account for the thermal energy of the twin clumps of plasma, $E_{\rm th}=1.6 \times 10^{50}\,$erg  (Sawada et al. 2011).  As illustrated in Figure 7, the lower relative velocity decreases the black hole mass.  

\begin{figure}[htbp]
\epsscale{0.8}
\plotone{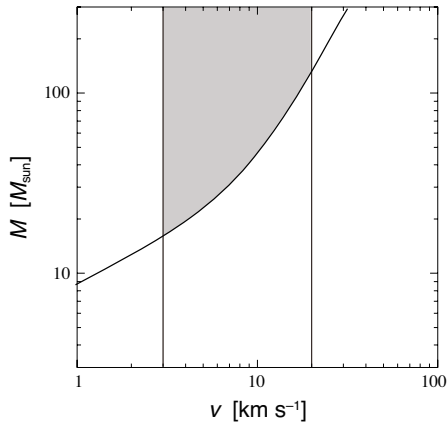}
\caption{Curve of $E_{\rm HL}=E_{\rm th}$ in the $M$-$v$ plane.  Gray shaded area indicates $E_{\rm HL}\geq E_{\rm th}$ in the permitted range of $v$.  }
\end{figure}

In order to confirm the above scenario, it is essential to find evidence for the cloud-to-cloud collision. For example, high-velocity wing emission, which bridges two molecular clouds with different velocities, was detected at the root of the pigtail molecular cloud (Matsumura et al. 2012).  Unfortunately, the $^{13}$CO data suffer from absorption features caused by local gas at $V_{\rm LSR}\sim 0\,{\rm km\ s^{-1}}$.  Observations in shock probes could detect broad-velocity emissions that connect the $-14\,{\rm km\ s^{-1}}$ and $+5\,{\rm km\ s^{-1}}$ clouds.  Bondi-Hoyle-Lyttleton accretion has been proposed for objects ranging from a young stellar object to an active galactic nucleus (Fromerth $\&$ Melia 2001; Throop $\&$ Bally 2008).  For example, Maeda et al. (2002) suggest that recent activity in the nucleus of our Galaxy could have been powered by the central black hole accreting material from an expanding supernova shock. Our scenario may add another example of this type of object.  In addition, we hypothesized an inactive compact object according to Sawada et al. (2011).  Direct evidence for the putative compact object must be acquired by X-ray or radio continuum observations in the future.

\section{Summary}
We observed the Tornado Nebula in the {\it J}=1--0 lines of CO, $^{13}$CO, and HCO$^+$ with the NRO 45-m telescope.  These observations and the reanalysis of the VLA archive data lead to the following conclusions.

\begin{enumerate}
\item{Extensive CO and $^{13}$CO {\it J}=1--0 maps delineated the distribution and kinematics of molecular gas in the direction of the Tornado Nebula. HCO$^+$ {\it J}=1--0 line emission was detected only in some limited regions.}
\item{The bulk of molecular gas traced by $^{13}$CO emission is mainly confined in two velocity components, the $-14\,{\rm km\ s^{-1}}$ and the $+5\,{\rm km\ s^{-1}}$ clouds.  These  $^{13}$CO clouds show clear spatial anti-correlation, suggesting physical interaction.}
\item{Spatially-extended OH 1720$\,$MHz emissions seem to be associated with the $-14\,{\rm km\ s^{-1}}$ and $+5\,{\rm km\ s^{-1}}$ clouds.  The spatial extension of OH 1720$\,$MHz emissions overlaps with the Tornado Nebula, indicating violent interaction between the Tornado Nebula and the $^{13}$CO clouds.}
\item{The LTE mass of the $-14\,{\rm km\ s^{-1}}$ and $+5\,{\rm km\ s^{-1}}$ clouds are $6 \times 10^5\, M_{\odot}$ and $1 \times 10^5\,M_{\odot}$, respectively.  The lower Virial mass/LTE mass ratio in the $+5\,{\rm km\ s^{-1}}$ cloud suggests that it is more tightly bound by self-gravity than the $-14\,{\rm km\ s^{-1}}$ cloud.}
\item{Gas with higher excitation temperature prefers the head of the Tornado Nebula, and the southern periphery of the $-14\,{\rm km\ s^{-1}}$ cloud.  Spatial correlation between the higher-excitation gas and the Tornado Nebula head lends more support for the interaction, especially with the $+5\,{\rm km\ s^{-1}}$ cloud.}
\item{Physical intimacy between the Tornado Nebula, the $-14\,{\rm km\ s^{-1}}$ cloud, and the $+5\,{\rm km\ s^{-1}}$ cloud suggests that these $^{13}$CO clouds might contribute to the formation of the Tornado Nebula.  We propose a formation scenario: (1) cloud-to-cloud collision generates a shock; (2) a high-density layer behind the shock passes a compact massive object; and (3) the Bondi-Hoyle-Lyttleton accretion onto the putative compact object activates it, ejecting bipolar jets to form the Tornado Nebula.}
 \end{enumerate}

\acknowledgments
We are grateful to the NRO staff for their excellent support of the 45-m observations.
The Nobeyama Radio Observatory is a branch of the National Astronomical Observatory of Japan, National Institutes of Natural Sciences.  We thank the anonymous referee for his/her constructive comments which improve this paper significantly.


\begin{thebibliography}{}
\bibitem[Becker and Helfand (1985)]{becker85} Becker, R. H., \& Helfand, D. J., 1985, \nat, 313, 115
\bibitem[Caswell et al. (1989)]{calwell89} Caswell, J. L., Kesteven, M. J., Bedding, T. R., \& Turtle, A. J., 1989, Proc. Astron. Soc. Australia, 8, 184
\bibitem[Castro et al. (2013)]{castro13} Castro, D., Slane, P., Carlton, A., \& Figueroa-Feliciano, E., 2013, \apj, 774, 36
\bibitem[Dickel et al. (1973)]{dickel73} Dickel, J. R., Milne, D. K., Kerr, A. R., \& Ables, J. G., 1973, Aust. J. Phys., 26, 379
\bibitem[Edger (2004)]{edgar04} Edgar, R., 2004, New Astronomy Reviews, 48, 843
\bibitem[Elitzur (1976)]{elitzur76} Elizur, M., 1976, \apj, 203, 124
\bibitem[Frail et al. (1996)]{frail96} Frail, D. A., Goss, W. M., Reynoso, E. M., Giacani, E. B., Green, A. J., \& Otrupcek, R. 1996, \aj, 111, 1651
\bibitem[Fromerth and Melia (2001)]{fromerth01} Fromerth, M. J., \& Melia, F., 2001, \apj, 549, 205
\bibitem[Helfand and Becker (1985)]{helfand85} Helfand, D. J., \& Becker, R. H., 1985, \nat 313, 118
\bibitem[Hewitt et al. (2009)]{hewitt09} Hewitt, J., Yusef-Zadeh, F., \& Wardle, M., 2009, \apj, 706, L270
\bibitem[Kutner and Ulich (1981)]{kutner81} Kutner, M. L., \& Ulich, B. L., 1981, \apj, 250, 341
\bibitem[Langer and Penzias (1990)]{langer90} Langer, W. D., Penzias, A. A., 1990, \apj, 357, 477
\bibitem[Lockett et al. (1999)]{lockett99} Lockett, P., Gauthier, E., \& Elitzur, M., 1999, \apj, 511, 235
\bibitem[Maeda et al. (2002)]{maeda02} Maeda, Y., Baganoff, F.K., Feigelson, E. D., Bautz, M. W., Brandt, W. N., Burrows, D. N., Doty, J. P., Garmire, G. P., Pravdo, S. H., Ricker, G. R., \& Townsley, L. K., 2002, \apj, 570, 671
\bibitem[Manchester (1987)]{manchester87} Manchester, R. N., 1987, \aap, 171, 205
\bibitem[Matsumura et al. (2012)]{matsumura12} Matsumura, S., Oka, T., Tanaka, K., Nagai, M., Kamegai, K., \& Hasegawa, T., 2012, \apj, 756, 87
\bibitem[Miley (1980)]{miley80} Miley, G., 1980, \araa, 18, 165
\bibitem[Oka et al. (1998)]{oka98} Oka, T., Hasegawa, T., Sato, F., Tsuboi, M., \& Miyazaki, A., 1998, \apjs, 118, 455
\bibitem[Oka et al. (2001)]{oka01} Oka, T., Hasegawa, T., Sato, F.,Tsuboi, M., Miyazaki, A., \& Sugimoto, M., 2001,\apj, 562, 348
\bibitem[Pollock (1985)]{pollock85} Pollock, A. M. T., 1985, \aap, 150, 339
\bibitem[Radhakrishnan et al. (1972)]{radhakrishnan72} Radhakrishnan, V., Goss, W. M., Murray, J. D., \& Brooks, J. W., 1972, \apjs, 24, 49
\bibitem[Sawada et al. (2011)]{sawada11} Sawada, M., Tsuru, G. T., Koyama, K., \& Oka, T., 2011, \apj, 63, 849
\bibitem[Sawada et al. (2008)]{sawada08} Sawada, T., Ikeda, N., Sunada, K., et al., 2008, \pasj, 60, 445
\bibitem[Shaber et al. (1985)]{shaver85} Shaver, P. A., Salter. C. J., Patnaik, A. R., van Gorkom, J. H., \& Hunt, G. C., 1985, \nat, 313, 113
\bibitem[Shull et al. (1989)]{shull89} Shull, J. M., Fesen, R. A., \& Saken, J. M., 1989, \apj, 346, 860
\bibitem[Slee et al. (1974)]{slee74} Slee, O. B., \& Dulk, G. A., 1974, in IAU Symp., 60, Galactic Radio Astronomy, ed. F. J. Kerr \& S. C. Simonson (Dordrecht: Reidel), 347
\bibitem[Sofue (2006)]{sofue06} Sofue, Y., 2006, \pasj, 58, 335
\bibitem[Solomon et al. (1987)]{solomon87} Solomon, P. M., Rivolo, A. R., Barrett, J., \& Yahil, A., 1987, \apj, 319, 730
\bibitem[Sunada et al. (2000)]{sunada00} Sunada, K., Yamaguchi, C., Nakai, N., et al., 2000, Proc. SPIE, 4015, 237
\bibitem[Throop and Bally (2008)]{throop08} Throop, H., B., \& Bally, J., 2008, \aj, 135, 2380
\bibitem[Wardle (1998)]{wardle98} Wardle, M., 1998, \mnras, 298, 507
\bibitem[Weiler and Panagia (1980)]{weiler80} Weiler, K., \& Panagia, N., 1980, \aap, 90, 269
\bibitem[Yusef-Zadeh et al. (1999)]{yusef99} Yusef-Zadeh, F, Goss, W. M., Roberts, D. A., Robinson, B., \& Frail, D. A., 1999, \apj, 527, 172 
\end{thebibliography}
\end{document}